\documentclass[aps,prl,preprint,superscriptaddress]{revtex4}

\usepackage{amssymb}
\usepackage{amsmath}
\usepackage{graphicx}

\usepackage{pdfpages} 

\newcommand{\beginsupplement}{%
        \setcounter{table}{0}
        \renewcommand{\thetable}{S\arabic{table}}%
        \setcounter{figure}{0}
        \renewcommand{\thefigure}{S\arabic{figure}}%
     }

\newcommand{\jeffHalf}{$j_{\mathrm{eff}}=1/2\ $}

\begin{document}

\title{Magnetization density distribution of Sr$_2$IrO$_4$: Deviation from a local \jeffHalf picture}

\author{Jaehong Jeong}
\email[]{hoho4@snu.ac.kr}
\affiliation{Universit\'e Paris-Saclay, CNRS, CEA, Laboratoire L{\'e}on Brillouin,  91191, Gif-sur-Yvette, France}
\affiliation{Center for Correlated Electron Systems, Institute for Basic Science (IBS), Seoul National University, Seoul 08826, Korea}

\author{Benjamin Lenz}
\affiliation{Centre de Physique Th\'eorique, Ecole Polytechnique, CNRS UMR7644, Institut Polytechnique de Paris, 91128 Palaiseau Cedex, France}
\affiliation{IMPMC , Sorbonne Universit{\'e}, CNRS , MNHN , IRD, 4 place Jussieu , 75252 Paris, France}

\author{Arsen Gukasov}
\affiliation{Universit\'e Paris-Saclay, CNRS, CEA, Laboratoire L{\'e}on Brillouin,  91191, Gif-sur-Yvette, France}

\author{Xavier Fabr\`eges}
\affiliation{Universit\'e Paris-Saclay, CNRS, CEA, Laboratoire L{\'e}on Brillouin,  91191, Gif-sur-Yvette, France}

\author{Andrew Sazonov}
\affiliation{Institute of Crystallography, RWTH Aachen University and J\"ulich Centre for Neutron Science (JCNS) at Heinz Maier-Leibnitz Zentrum (MLZ), 85747 Garching, Germany}

\author{Vladimir Hutanu}
\affiliation{Institute of Crystallography, RWTH Aachen University and J\"ulich Centre for Neutron Science (JCNS) at Heinz Maier-Leibnitz Zentrum (MLZ), 85747 Garching, Germany}

\author{Alex Louat}
\affiliation{Laboratoire de Physique des Solides, Universit\'e Paris-Sud, UMR 8502, 91405 Orsay, France}

\author{Dalila Bounoua}
\affiliation{Universit\'e Paris-Saclay, CNRS, CEA, Laboratoire L{\'e}on Brillouin,  91191, Gif-sur-Yvette, France}

\author{Cyril Martins}
\affiliation{Laboratoire de Chimie et Physique Quantiques, UMR 5626, Universit\'e Paul Sabatier, 118 route de Narbonne, 31400 Toulouse, France}

\author{Silke Biermann}
\affiliation{Centre de Physique Th\'eorique, Ecole Polytechnique, CNRS UMR7644, Institut Polytechnique de Paris, 91128 Palaiseau Cedex, France}
\affiliation{Coll\`ege de France, 11 place Marcelin Berthelot, 75005 Paris, France}
\affiliation{Department of Physics, Division of Mathematical Physics, Lund University, Professorsgatan 1, 22363 Lund, Sweden}
\affiliation{European Theoretical Spectroscopy Facility, 91128 Palaiseau, France, Europe}

\author{V\'eronique Brouet}
\affiliation{Laboratoire de Physique des Solides, Universit\'e Paris-Sud, UMR 8502, 91405 Orsay, France}

\author{Yvan Sidis}
\affiliation{Universit\'e Paris-Saclay, CNRS, CEA, Laboratoire L{\'e}on Brillouin,  91191, Gif-sur-Yvette, France}

\author{Philippe Bourges}
\email[]{philippe.bourges@cea.fr}
\affiliation{Universit\'e Paris-Saclay, CNRS, CEA, Laboratoire L{\'e}on Brillouin,  91191, Gif-sur-Yvette, France}

\begin{abstract}

$5d$ iridium oxides are of huge interest due to the potential for new quantum states driven by strong spin-orbit coupling.
The strontium iridate Sr$_2$IrO$_4$ is particularly in the spotlight because of  the so-called $j_\text{eff}=1/2$ state  consisting of a quantum superposition of the three local $t_{2g}$ orbitals with  -- in its most simple version --
nearly equal population, which stabilizes an unconventional Mott insulating state. Here, we report an anisotropic and aspherical magnetization density distribution measured by polarized neutron diffraction in a magnetic field up to 5~T at 4~K, which strongly deviates from a local \jeffHalf picture
even when distortion-induced deviations from the equal  weights of the orbital  populations are taken into account.
Once reconstructed by the maximum entropy method  and multipole expansion model refinement, the magnetization density  shows cross-shaped positive four lobes along the crystallographic tetragonal axes with a large spatial extent, showing that the $xy$ orbital contribution is dominant. 
The analogy to the superconducting copper oxide systems might then be weaker than commonly thought.

 %

\end{abstract}

\maketitle

\begin{figure*}
	\includegraphics[width=0.8\textwidth,clip]{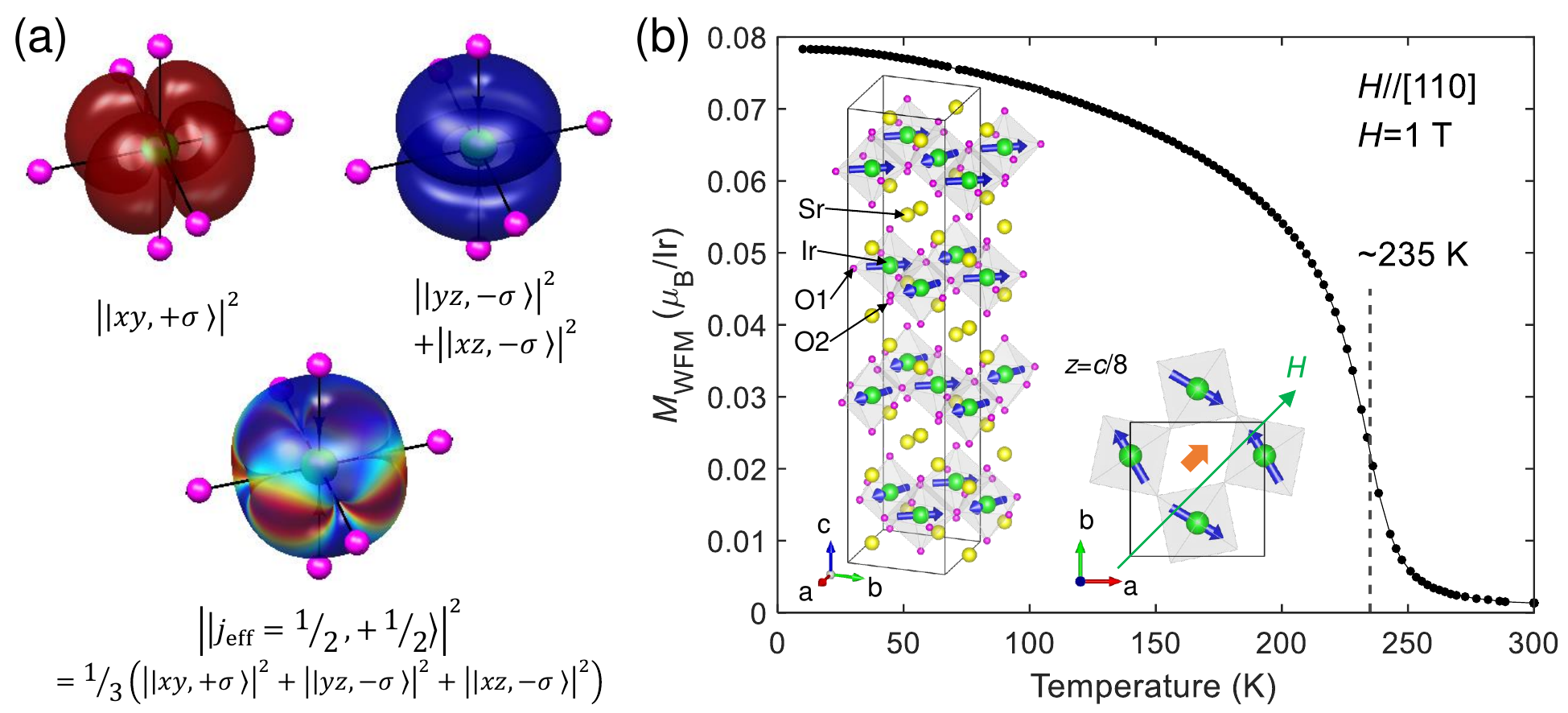}
	\caption{\textbf{The $j_\text{eff}=1/2$ states and uniform magnetization of Sr$_2$IrO$_4$.}
	\label{fig:jeff}\textbf{(a)} Illustration of the electron and spin density distributions for the ideal (distorted) $j_\text{eff}=1/2, m_j=1/2$ state, which consists of three (nearly)  equally populated $t_{2g}$ orbitals with mixed spin states.
	The red and blue colors represent spin-up and spin-down states, respectively.
	\label{fig:mag}\textbf{(b)} The magnetization vs temperature curve under $H=1$~T ($H//[110]$). 
	It exhibits a weak ferromagnetic moment inherited from the AF-II order transition~\cite{ye2013} at $\approx$235~K. The inset shows the crystal and magnetic structure of Sr$_2$IrO$_4$ for an applied magnetic field along $H//[110]$.}
\end{figure*}

Sr$_2$IrO$_4$ possesses a tetragonal structure with $I4_1/acd$ space group, in which the IrO$_6$ octahedra are rotated by $\approx$11$^\circ$ around the $c$-axis with an opposite phase for the neighboring Ir ions and it orders antiferromagnetically below $T_\text{N}\approx230$~K~\cite{huang1994,kim2008,ye2013,dhital2013,sung2016}.
Strong spin-orbit coupling (SOC) stabilizes an unconventional Mott insulating ground state, which is commonly described by a spin-orbital product state within a so-called $j_\text{eff}=1/2$ model~\cite{kim2008,kim2009,jackeli2009,wang2011}.
In  the most simple version of this model, $5d$ electrons at the Ir$^{4+}$~($5d^5$) ions occupy the $t_{2g}$ states with an effective angular momentum $l_\text{eff}=1$, which are split by the relatively large SOC into a $j_\text{eff}=1/2$ doublet and a $j_\text{eff}=3/2$ quartet.
The Coulomb repulsion induces a gap in the narrow half-filled $j_\text{eff}=1/2$ band, and stabilizes the Mott insulating state with the pseudospin $j_\text{eff}=1/2$~\cite{kim2008,jackeli2009}, which consists of three equally populated spin-orbital components in the $t_{2g}$ band (Fig.~\ref{fig:jeff}a):
\begin{equation}
	\left|j_\text{eff}=\frac{1}{2}, \pm\frac{1}{2}\right> = \frac{1}{\sqrt{3}} \left( \left|xy,\pm\sigma\right> \pm
		\left|yz,\mp\sigma\right> + i\left|xz,\mp\sigma\right> \right).
\label{Eq:1}
\end{equation}
where  $\pm\sigma$ denotes the spin of the electrons.

While resonant and inelastic X-ray scattering~\cite{kim2009,kim2012} gave credit to a description in terms of these $j_\text{eff}=1/2$ states \cite{bertinshaw2019}, the simple description with equal weights for the orbital populations
has been questioned owing to the tetragonal distortion that is not negligible~\cite{chapon2011,haskel2012,morettisala2014}.
Indeed, the most simple model with equal weights is realized only for a perfect cubic symmetry, while
any lattice distortions (compression, elongation or tilting of the octahedra around the $c$-axis) split the $t_{2g}$ orbitals into three non-degenerate Kramers doublets \cite{abragam2012,jackeli2009,chapon2011,perkins2014,morettisala2014,lenz2019}, driving the lowest energy state away from the equal weight case of Eq. (\ref{Eq:1}). In the following we adopt the nomenclature common to the field and still refer to  the states that diagonalize the local Hamiltonian in the presence of the distortions present in Sr$_2$IrO$_4$  as $j_\text{eff}=1/2$ and $j_\text{eff}=3/2$ states.  Fig. 1a compares the electron and spin density distributions  corresponding to the perfectly cubic and the distorted cases.

In addition, a strong hybridization between Ir $5d$ and O $2p$ orbitals, which seems to be natural for a large spatial extent of $5d$ orbitals, has been proposed to account for a large reduction of the ordered magnetic moment~\cite{kim2008} as well as for antiferromagnetic (AFM) exchange interactions between the nearest-neighboring Ir ions and for the canted magnetic moments following the octahedral rotations~\cite{jackeli2009,perkins2014}. 
The strong hybridization of the $d$-orbitals with the $p$-orbitals of the ligand oxygen is reminiscent of K$_2$IrCl$_6$~\cite{lynn1976} and the isostructural ruthenate Ca$_{1.5}$Sr$_{0.5}$RuO$_4$~\cite{gukasov2002}, where similar covalency effects have been reported.
In Sr$_2$IrO$_4$, recent muon spin relaxation measurements have suggested the formation of oxygen moments~\cite{miyazaki2015}, and charge redistribution between adjacent IrO$_2$  and SrO layers has been revealed using electron spin resonance measurement~\cite{bogdanov2015}. 
Further, unusual magnetic multipoles have been proposed to be observed by neutron diffraction~\cite{lovesey2014} and recently a hidden magnetic order having the same symmetry as a loop-current state has been observed by polarized neutron diffraction~\cite{jeong2017}. 

\begin{figure*}
	 \includegraphics[width=0.9\textwidth,clip]{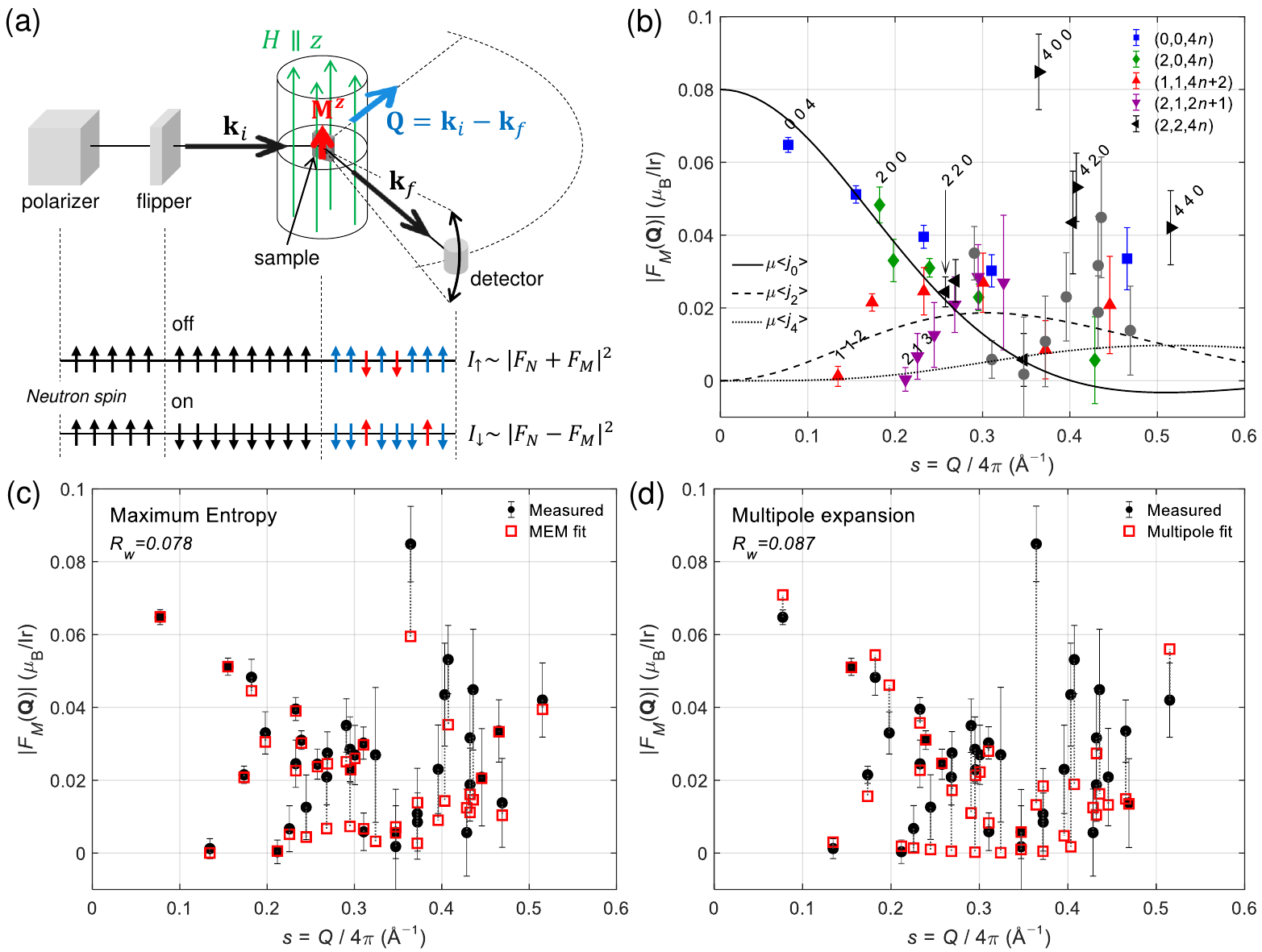}
	\caption{\textbf{Polarized neutron diffraction setup and measured neutron magnetic structure factor of Sr$_2$IrO$_4$.}
	\label{fig:setup}\textbf{(a)} The experimental setup for a polarized neutron diffraction experiment. The arrows at the bottom denote a spin polarization of neutrons. The vertical direction corresponds to either the 
$[010]$ or $[\bar{1}10]$ crystallographic direction for each sample orientation \cite{suppall}.
		\label{fig:formfactor}\textbf{(b)} The magnetic structure factor of all measured momentum transfer $Q$ with the theoretical radial integrals $\left<j_n\right>$ for isolated Ir$^{4+}$ ions. A  series of reflections along the $(0,0,l)$ are highlighted: $(0,0,4n)$ in blue squares, $(2,0,4n)$ in green diamonds, $(1,1,4n+2)$ in red up-triangles, $(2,1,2n+1)$ in purple down-triangles, and $(2,2,4n)$ in black left-triangles. The $(4,0,0)$, $(4,2,0)$ and $(4,4,0)$ are also presented in black right-triangles, and the rest in grey circles.  Measured and fitted magnetic structure factors  $\left|F_M(\mathbf{Q})\right|$ for  \textbf{(c)} the optimized MEM result and  \textbf{(d)} optimized multipole expansion result. }
\end{figure*}

The magnetic moments of Ir ions are confined in the $ab$-plane and track the staggered octahedral rotation in an $-++-$ sequence along the $c$-axis in the unit cell~\cite{ye2013}.  
Owing to this canted AFM structure, each IrO$_2$ layer has a weak ferromagnetic (WFM) moment along the principal crystallographic axis in the $ab$-plane at zero magnetic field.
This WFM is compensated due to  the $-++-$ stacking sequence whereas, in a magnetic field higher than $H_c\approx0.3$~T~applied in the $ab$-plane~\cite{kim2008,ye2013}, a net homogeneous WFM moment appears in the plane  (inset of Fig.~\ref{fig:mag}b) above the metamagnetic transition.  
Remarkably, this WFM moment follows the direction of applied magnetic field in the $ab$-plane~\cite{fruchter2016,nauman2017,porras2019} and attains a saturation value of $\approx$0.08$\mu_\text{B}$/Ir in the field of 1~T~\cite{fruchter2016}. In the current experiment, a uniform magnetic field ($H$) up to 5~T has been applied along the vertical direction (Fig.~\ref{fig:setup}a).  
The IrO$_6$ octahedral rotation generates two additional magnetic terms in the simple Heisenberg-type Hamiltonian \cite{porras2019}: $J_z$ and Dzyaloshinskii-Moriya terms,  which restrict the angle between adjacent pseudospins to $\pi+2\alpha$ with the octahedral rotation angle $\alpha$~\cite{jackeli2009}. However, it does not break the in-plane rotational symmetry as the pseudospins are free to rotate in the plane while keeping the same canting angle between them  (the situation is shown in the inset of  Fig. \ref{fig:jeff}b for a field applied along the $[110]$ direction). 
Therefore, under the applied magnetic field, the WFM moment does not interlock with the rotation of IrO$_6$ octahedra in contrast with the AFM staggered moment at zero field. 

The existence of this WFM allows us to probe the magnetization density distribution in crystals by polarized neutron diffraction (PND).  This technique is unique because it provides direct information about the 3-dimensional distribution of the magnetization throughout the unit cell, which in turn allows for a determination of the symmetry of occupied orbitals. This method has been  successfully used in the study of ferromagnetic ruthenate Ca$_{1.5}$Sr$_{0.5}$RuO$_4$, isostructural to Sr$_2$IrO$_4$, where an anomalously high spin density at the oxygen site and the $xy$ character of the Ru $d$-orbitals have been reported~\cite{gukasov2002}.

The typical experimental setup for PND, shown in Fig.~\ref{fig:setup}a,  consists of a neutron polarizer, a flipping device that reverses the incident neutron polarization, a magnet and a detector.
The sample is magnetized by a magnetic field applied along the vertical axis and scattering intensities of Bragg reflections for the two opposite states (spin-up and spin-down) of the incident polarization are measured. 
They are used to calculate the so-called flipping ratio, allowing access to the Fourier components of the magnetization density, as
\begin{equation}
	R_\text{PND}=\frac{I_\uparrow}{I_\downarrow}=\frac{F_N^2+2p \sin^2\alpha F_N F_M + \sin^2\alpha {F_M}^2} {F_N^2 - 2 p e \sin^2\alpha F_N F_M + \sin^2\alpha {F_M}^2},
	\label{eq:FR}
\end{equation}
where $F_N$ is the nuclear structure factor and $F_M$ is the magnetic structure factor.
$p$ and $e$ are the polarization efficiency of the polarizer and flipper, respectively, and $\alpha$ is the angle between the scattering vector and the magnetization~\cite{suppall}.
The flipping ratios $R_\text{PND}$ of more than 280 $(hkl)$ reflections were measured in the weakly ferromagnetic state above the metamagnetic transition at 2~K for two magnetic field orientations, $H\|[010]$ and $H\|[\bar{1}10]$  (well above the critical field  $H_c\approx0.3$~T~\cite{kim2009,porras2019}). 
The measured intensities for two orientations were averaged~\cite{suppall}.
As shown in Fig.~\ref{fig:formfactor}b, the magnetic structure factors $F_M$ were directly obtained from the measured flipping ratios by using Eq.~(\ref{eq:FR}) and known nuclear structure factors $F_N$.
For convenience, the amplitudes are given in Bohr magnetons, normalized by the number of Ir atoms (8) in the unit cell, and taken in absolute values to remove alternating signs of the phase factor. 
The amplitude, $F_M(0)$, is imposed in agreement with the saturation moment (0.08$\mu_\text{B}$/Ir) given by the uniform magnetization measurement~\cite{fruchter2016}. 

 In the \textit{dipole approximation}, {\it i.e.} at low momentum transfer, $F_M(Q)$ is usually described by a smooth decreasing function of $Q$, called the magnetic form factor, corresponding to a linear combination of radial integrals calculated from the electronic radial wave function. Instead in Fig.~\ref{fig:formfactor}b, one observes a large distribution of the measured structure factor indicating unusually large anisotropy. That large anisotropy is explained by a predominance of $xy$-orbital as shown below using the reconstruction of the magnetization density in real space.  The theoretical radial integrals $\left<j_n\right>$ for an isolated Ir$^{4+}$ ion \cite{kobayashi2011} are also shown in Fig.~\ref{fig:formfactor}b for comparison. 
We recall that $\left<j_2\right>$,$\left<j_4\right>$ and higher-order integrals are needed to describe the departures from spherical symmetry. 
As seen from Fig.~\ref{fig:formfactor} except for the $(0,0,l)$ reflections, decreasing gradually with increasing $Q$, the majority of reflections strongly deviate from any expected smooth curve.
Moreover, while the $(0,0,4n)$, $(2,0,4n)$ and $(2,2,4n)$ reflections are close to the $\left<j_0\right>$ curve in a small $Q$ region,  the $(1,1,4n+2)$ and $(2,1,2n+1)$ reflections deviate from it quite strongly.
This indicates an aspherical magnetization density, which is typical of ions with one or two unpaired electrons in the $d$-orbitals~\cite{lynn1976,shamoto1993,zaliznyak2004}.
In addition, one can see that high-$Q$ reflections like the $(4,0,0)$, $(4,2,0)$ and $(4,4,0)$ ones show anomalously large values. 

Next, a real space visualization has been performed by a reconstruction of the magnetization density, using two different very well-established and widely used approaches; a model-free maximum entropy method (MEM)~\cite{papoular1990} and a quantitative refinement using the multipole expansion of the density function~\cite{coppens1997}. Both techniques have advantages and limits and should be employed where they are the most efficient. Typically, no assumption is made for the initial magnetization distribution in MEM whereas the $d$-orbitals shape is constrained in the multipole expansion.  

\begin{figure*}
      \includegraphics[width=\textwidth,clip]{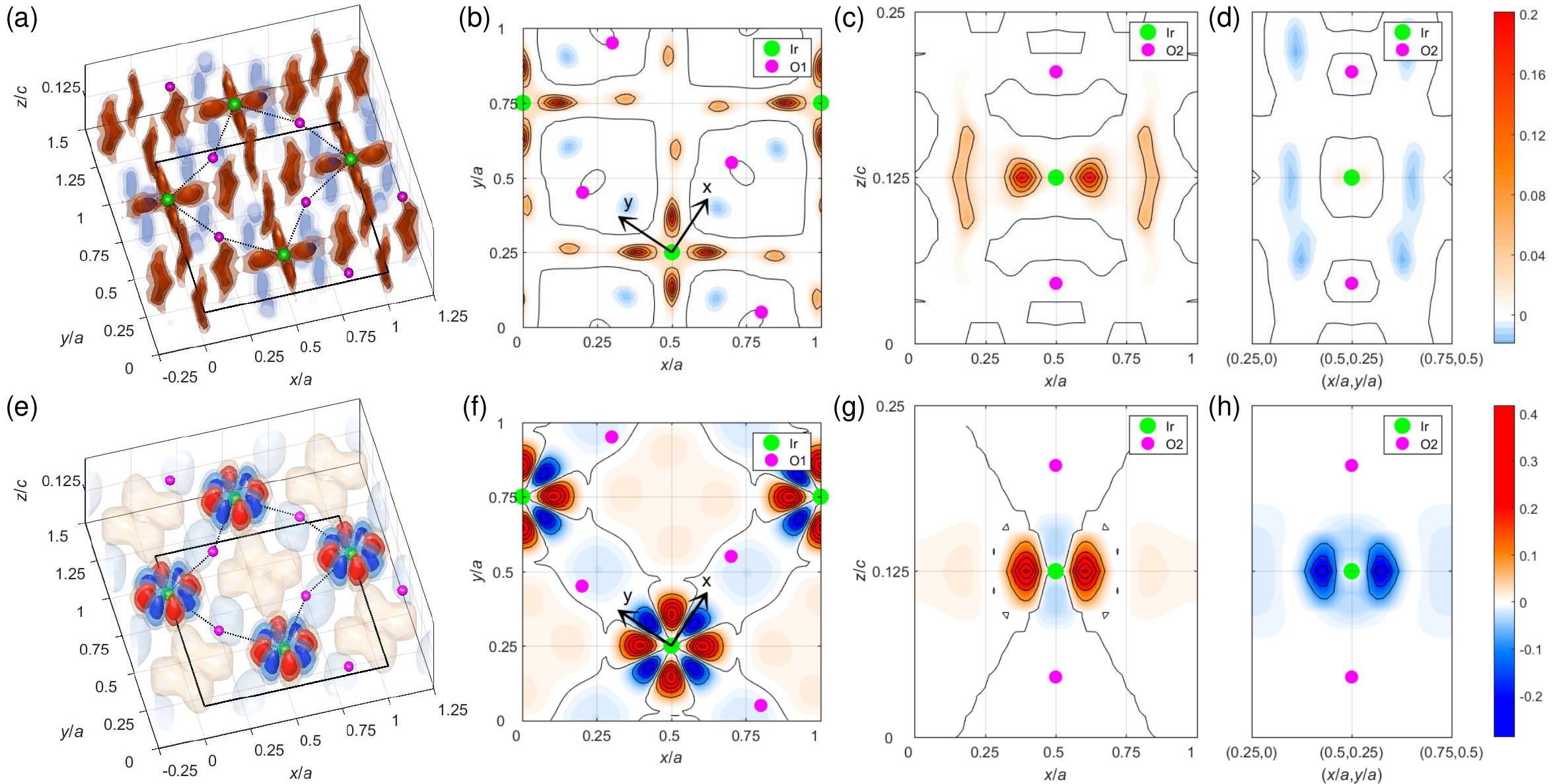}
	\caption{\textbf{Magnetization density distribution reconstructed by MEM and multipole expansion refinement.}
	\label{fig:magden} 3D magnetization density distribution on the $z=c/8$ layer reconstructed by \textbf{(a)} the MEM and \textbf{(e)} multipole expansion model refinement. Isosurfaces encompassing 30\%, 50\% and 70\% of the volume density are plotted with a desceding opacitiy according to their isovalues. Red and blue surfaces denote positive and negative magnetizations, repectively. The solid square and dotted lines denote the unit cell and Ir-O bonds, respectively.
	 Sliced density contour maps at \textbf{(b,f)} $(x,y,c/8)$, \textbf{(c,g)} $(x,a/4,z)$ and \textbf{(d,h)} $(a/4+x,x,z)$ are also shown for both methods. The contour step is 0.04 and 0.08$\mu_\text{B}/$\AA$^3$ for \textbf{(b-d)} and \textbf{(f-h)}, respectively.  The black arrows correspond to the Ir-O bonding directions.}
\end{figure*}

Since the crystal structure is centrosymmetric, the magnetization density can be directly reconstructed from the measured magnetic structure factors by MEM~\cite{papoular1990}. 
Fig.~\ref{fig:magden}a-d, shows the 3-dimensional magnetization density reconstructed by using a conventional flat density prior.  A positive magnetization density in red color denotes a magnetic moment density parallel to the applied magnetic field and a negative one in blue is antiparallel. There are three key features to be noted in the figure.
First, the magnetization density at Ir sites has four positive density lobes directed along the $a$, $b$ axes, corresponding to a dominant positive magnetization density of $d_{xy}$ orbital symmetry (Fig.~\ref{fig:magden}b). The two other components of the effective $j_\text{eff}=1/2$ state model, $d_{yz}$ and $d_{xz}$, which would form an axially symmetric doughnut-shaped density 
above and below the $xy$ plane (see Fig.~\ref{fig:jeff}a), does not appear as seen in Fig.~\ref{fig:magden}c,d. 
Thus the WFM density originates predominantly from the $xy$ orbital (a schematic illustration of the magnetic components in this situation is given in the Supplemental file (Fig. S4) \cite{suppall}, in contrast with the local  \jeffHalf picture). Second, positive density lobes are very strongly elongated, in such a way that some magnetization density is delocalized well beyond of the IrO$_6$ octahedra. 
Third, contrary to the expectation of strong iridium oxygen ligand hybridization, no visible induced magnetization density appears at the oxygen sites.
Actually, no significative polarization dependence has been found in any of dozens measured  $(2,1,2n+1)$ reflections  where oxygen atoms contribute. This is in contrast with the isostructural $4d$ compound Ca$_{1.5}$Sr$_{0.5}$RuO$_4$, where $\sim$20\% of the magnetic moment is transferred to the in-plane O sites~\cite{gukasov2002}. 
However, one can notice the presence of a negative magnetic density mostly along the Ir-O direction existing between the large positive lobes.
In fact, a significant negative magnetization density as large as half of the net moment is essential for a better description in the MEM analysis~\cite{suppall}.

To confirm the symmetry found by MEM,  multipole expansion was performed for an alternative refinement of the WFM density~\cite{hansen1978,coppens1997}. It is composed of radial and angular parts: Slater-type radial wave functions and real spherical harmonic density functions~\cite{suppall}. In Fig.~\ref{fig:magden}e-h, the magnetization density distribution with the best refinement is shown.
The main positive magnetization density lobes located between the local $x$- and $y$-axis appear clearly, which corresponds to the $d_{xy}$ symmetry.  
Therefore, the multipole expansion model fully confirms the  $d_{xy}$ symmetry found by MEM. A benefit of the multipole method is to determine
the contribution of all five $d$-orbitals to the magnetization.  Using the orbital-multipole relations~\cite{coppens1997}, the magnetic moments on each orbitals were obtained as: $+0.48$, $-0.051$, $-0.035$ and $-0.314\mu_\text{B}$/Ir for $d_{xy}$, $d_{yz/xz}$, $d_{z^2}$ and $d_{x^2-y^2}$, respectively.
Thus a positive $d_{xy}$ and to a lesser extent a negative $d_{x^2-y^2}$ orbital are dominant in the refinement  (the latter  effect is minor in the MEM method), while the $d_{yz/xz}$ orbitals are barely populated. Interestingly, the admixture of $d_{x^2-y^2}$ character to the \jeffHalf orbital also has been found in first principles simulations~\cite{martins2017}. 

 It is obvious that the refinement of multipoles with a single radial exponent cannot fit the widely delocalized density. 
Therefore, we introduce in the refinement a second radial exponent to describe the delocalized Ir density.
Such a model shows a considerably better agreement factor ($R_w\sim0.09$) compared to the model with a single radial exponent ($R_w\sim0.18$) (see Supplemental section V\cite{suppall}).
It confirms the anomalously large spatial extent of the magnetization density of Ir found by the MEM analysis. To appreciate the relevance of the obtained magnetization maps, we calculate the magnetic structure factors from the optimized MEM and multipoles results.  By plotting them along with the measured ones in  Fig.~\ref{fig:formfactor}c (for MEM) and Fig.~\ref{fig:formfactor}d (for multipoles), one sees that the calculated densities with MEM reproduce better the experimental data.

While the standard modeling of layered iridates by means of an anisotropic super-exchange Hamiltonian within the \textit{effective local} $j_{\mathrm{eff}}=1/2$ picture correctly captures the WFM moment of the ground state and explains the most salient magnetic properties of Sr$_2$IrO$_4$ \cite{jackeli2009,perkins2014,martins2018,bertinshaw2019,lenz2019},
this local $j_{\mathrm{eff}} = 1/2$ model is at odds with the present findings:
its simplified version with equal orbital weights would suggest a
homogeneous magnetization density, while taking into account distortions
would -- even worse -- enhance the $xz$ and $yz$ orbital weight of the hole
and thus of the WFM (see Fig.~\ref{fig:jeff}a).

These considerations give an additional twist to the exotic properties of Sr${}_2$IrO${}_4$ and the possibilities of modeling them as well as to the relationship to superconducting copper oxides.  Recently,  our PND results has been interpreted  in terms of spin anapole \cite{lovesey2019}, pointing towards the existence of multipole correlations that goes beyond the local $j_{\mathrm{eff}} = 1/2$ picture.  An alternative interpretation based on a momentum-dependent composition of the orbital carrying the hole in terms of atomic $t_{2g}$ states will be published elsewhere  \cite{lenz2020}.
In this type of model,  the hole resides in an orbital that results from a non-local, that is a $\mathbf{k}$-dependent superposition of Wannier functions of  $t_{2g}$ character.  In this light, it is less surprising that neither the simplified \jeffHalf picture discussed above nor the state
that takes into account the structural distortions but
remains restricted to a local superposition of atomic orbitals
describes our present experimental findings.


In summary, using PND we have evidenced a magnetization density distribution in Sr$_2$IrO$_4$ that is inconsistent with the local \jeffHalf picture. 
The measured magnetic structure factor shows a strong axial anisotropy and anomalous values at large $Q$, which indicate an aspherical magnetization density distribution with a significant orbital contribution. Real space visualization exhibits a dominant $d_{xy}$ orbital character with highly elongated lobes of Ir magnetization densities towards the next Ir atoms. Although a strong $d$-$p$ hybridization is expected in Sr$_2$IrO$_4$, the magnetization density at the ligand oxygen sites is barely present. Our results elucidate that the ground state of Sr$_2$IrO$_4$ substantially deviates from the commonly accepted local \jeffHalf state.

\begin{acknowledgments}
We acknowledge supports from the projects NirvAna (contract ANR-14-OHRI-0010) and SOCRATE (ANR-15-CE30-0009-01) of the French Agence Nationale de la Recherche (ANR), by the Investissement d’Avenir LabEx PALM (GrantNo. ANR-10-LABX-0039-PALM) and by the European Research Council under grant agreement CorrelMat-617196 for financial support.  
J.J. was supported by an Incoming CEA fellowship from the CEA-Enhanced Eurotalents program, co-funded by FP7 Marie-Sklodowska-Curie COFUND program (Grant Agreement 600382) and by the Institute of Basic Science (IBS) in Korea (Grant No. IBS-R009-G1).  Instrument POLI at Maier-Leibnitz Zentrum (MLZ) Garching is operated in cooperation between RWTH Aachen University and Forschungszentrum J\"{u}lich GmbH (J\"{u}lich-Aachen Research Alliance JARA). We thank Pr S.V. Lovesey for valuable comments on the manuscript and J. Porras for scientific discussions. 

\end{acknowledgments}

\bibliography{bib_Sr2IrO4}

\clearpage

\section{Supplementary Material}
 \beginsupplement


\section*{Supplementary material}

# Supplemental Material

# Magnetization density distribution in Sr$_2$IrO$_4$: Deviation from a local $j_{\text{eff}} = 1/2$ picture

Jaehong Jeong,[1,2,*] Benjamin Lenz,[3,4] Arsen Gukasov,[1] Xavier Fabrèges,[1] Andrew Sazonov,[5] Vladimir Hutanu,[5] Alex Louat,[6] Dalila Bounoua,[1] Cyril Martins,[7] Silke Biermann,[3,8,9,10] Véronique Brouet,[6] Yvan Sidis,[1] and Philippe Bourges[1,†]

[1] Université Paris-Saclay, CNRS, CEA,
Laboratoire Léon Brillouin, 91191, Gif-sur-Yvette, France

[2] Center for Correlated Electron Systems, Institute for Basic Science (IBS),
Seoul National University, Seoul 08826, Korea

[3] Centre de Physique Théorique, Ecole Polytechnique, CNRS UMR7644,
Institut Polytechnique de Paris, 91128 Palaiseau Cedex, France

[4] IMPMC, Sorbonne Université, CNRS, MNHN,
IRD, 4 place Jussieu, 75252 Paris, France

[5] Institute of Crystallography, RWTH Aachen University and Jülich Centre for Neutron
Science (JCNS) at Heinz Maier-Leibnitz Zentrum (MLZ), 85747 Garching, Germany

[6] Laboratoire de Physique des Solides,
Université Paris-Sud, UMR 8502, 91405 Orsay, France

[7] Laboratoire de Chimie et Physique Quantiques,
UMR 5626, Université Paul Sabatier,
118 route de Narbonne, 31400 Toulouse, France

[8] Collège de France, 11 place Marcelin Berthelot, 75005 Paris, France

[9] Department of Physics, Division of Mathematical Physics,
Lund University, Professorsgatan 1, 22363 Lund, Sweden

[10] European Theoretical Spectroscopy Facility, 91128 Palaiseau, France, Europe

## I. SAMPLE PREPARATION AND CHARACTERIZATION

High-quality single crystals of $Sr_2IrO_4$ were grown at Laboratoire de Physique des Solides at Orsay using a self-flux method in the platinum crucibles. The temperature dependence of the uniform magnetization was measured using a SQUID under a magnetic field of $H = 1$ T along the [110] direction (Fig. **S1**a). $Sr_2IrO_4$ undergoes a canted antiferromagnetic (AFM) transition. The ordered AFM moment is found to be about 0.21 $\mu_B/Ir$[1] or about 0.36 $\mu_B/Ir$[2] by single crystal neutron diffraction. We here discussed the weak in-plane ferromagnetic moment due to the moment canting (see Fig 1.b in the main text). In presence of the magnetic field above $H > 0.2$ T, the system exhibits a ferromagnetic component at $T \approx 235$ K, where the derivative $dM/dT$ diverges. The saturation weak ferromagnetic (WFM) moment is $\approx 0.08\mu_B/Ir$ at $T = 10$ K. It is well consistent with the reported value for single crystals grown under the best conditions[3]. For the polarized neutron diffraction measurements, which require a large amount of the sample, 50 single crystals of a small size (typically $< 2 \times 1 \times 0.2$ mm$^3$) were coaligned within a diameter of 10 mm on an aluminum plate (Fig. **S1**b).

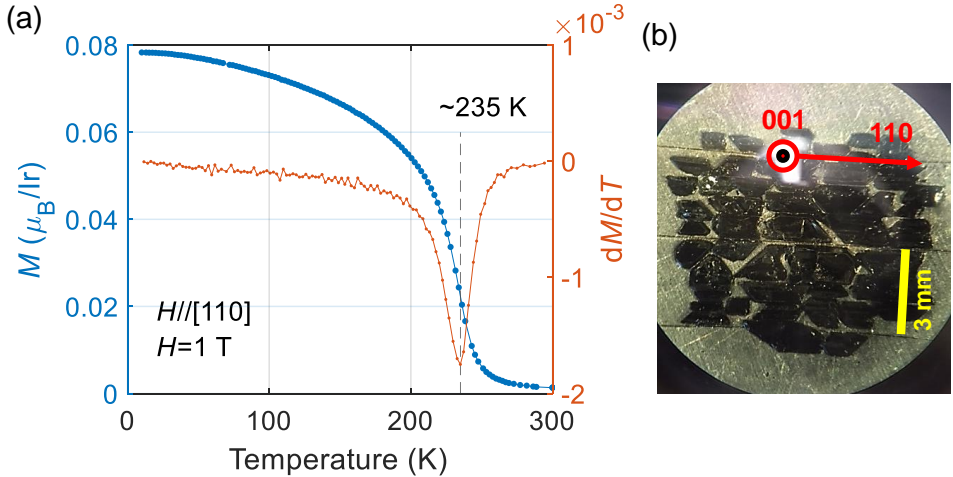

FIG. **S1**. (a) Temperature dependence of the uniform magnetization and its derivative measured by a SQUID. (b) Coaligned 50 single crystals on an aluminum plate.

## II. POLARIZED NEUTRON DIFFRACTION

Polarized neutron diffraction (PND) measurements were performed on three different polarized neutron diffractometers, *5C1* and *6T2* at the *ORPHÉE* reactor (LLB CEA Saclay)[4,5], and *POLI* at *FRM-II* (Maier-Leibnitz Zentrum Garching). Thermal and hot polarized neutron wavelengths were used with $\lambda$=1.4 Å (6T2), 1.15 Å (POLI) and 0.84 Å (5C1). A vertical uniform magnetic field of 5 T on 5C1/6T2 at LLB and 2.2 T on POLI at FRMII has been applied at the sample position. As the weak ferromagnetic component saturates above $\sim$ 1 T, that difference between the applied fields is not significant. The temperature was set at 4 K in the antiferromagnetic state of $Sr_2IrO_4$.

Incoming neutrons were polarized by a Heusler alloy monochromator, a supermirror bender or a $^3$He filter. The direction of incident neutron polarization – spin-up or spin-down – was chosen by a polarization flipper. The incident spin-polarized neutrons are scattered by the nuclei and by the local magnetization, which are denoted by nuclear and magnetic scattering, respectively. The total neutron cross section for a non-chiral system is given by

$$\sigma = F_N F_N^* + \mathbf{M}_\perp \cdot \mathbf{M}_\perp^* + \mathbf{P}_i \cdot (\mathbf{M}_\perp F_N^* + \mathbf{M}_\perp^* F_N), \tag{S1}$$

where $F_N$ is the nuclear structure factor, $\mathbf{M}_\perp$ is the magnetic interaction vector and $\mathbf{P}_i$ is the incident neutron polarization vector. The magnetic interaction vector is given as $\mathbf{M}_\perp = \hat{Q} \times F_M(\mathbf{Q}) \times \hat{Q}$ with the scattering vector $\mathbf{Q}$ and $F_M(\mathbf{Q})$ the magnetic structure factor, *i.e.*, the Fourier transform of the magnetic moment distribution. Therefore, only magnetic components perpendicular to the scattering vector participate in the magnetic scattering. The last *magnetic-nuclear interference* term appears only for a polarized beam ($|P_i| > 0$), and it contributes to the scattering for a mixed nuclear-magnetic Bragg reflection with propagation vector $\mathbf{q} = 0$. It corresponds to the magnetic response at the Brillouin zone center, $\Gamma$, point that would differ from the magnetic response at the aniferromagnetic propagation wave vector. In all the manuscript, we are only speaking about the weak ferromagnetic component of the moment. This method is well established for paramagnetic and ferromagnetic (FM) systems. It gave access, e.g., to the 3$d$-orbital population in ferromagnetic insulator $YTiO_3$[6]. While in conventional spin density studies either a positive or a negative spin component can be present at a given ion, in the iridates thanks to the spin-orbit coupling both positive and negative densities can coexist at the same Ir site (Fig. 1a of the manuscript)[7]. To access this intra-atomic variation of magnetization density high

resolution polarized hot neutron diffraction data are needed. Here we have performed PND (i) to establish the symmetry of the Ir $5d$ orbitals occupied by unpaired electrons and (ii) to check the presence of unpaired electron density on the oxygen ligand.

In case of PND where incoming neutron polarization is directed parallel or antiparallel to an external magnetic field, one measures the intensity of the scattered neutrons for the two states of the flipper. We note that no final polarization analysis is performed. Therefore, the measured intensity, obtained for each state of the flipper, is written as a function of the real structure factors,

$$I_\uparrow = F_N^2 + M_\perp{}^2 + 2P_i F_N M_\perp^z$$
$$I_\downarrow = F_N^2 + M_\perp{}^2 - 2P_i' F_N M_\perp^z,$$
(S2)

where $M_\perp$ is now the projection of the magnetic interaction vector $\mathbf{M}_\perp$ along the polarization vector $\mathbf{P}_i$ with $P_i$ and $P_i'$ being the incident polarization for spin-up and spin-down, respectively. The PND measurement is thus only sensitive to the uniform magnetization along the applied magnetic field, which is the vertical $z$-direction in our case. Using $P_i = p$, $P_i' = pe$, $M_\perp = \sin \alpha F_M$ and $M_\perp^z = \sin^2 \alpha F_M$ where $\alpha$ is the angle between the magnetic interaction vector and the scattering vector, the flipping ratio is given as Eq. (2) in the main text with the polarization efficiency $p$ of the polarizer and the flipping efficiency $e$ of the flipper. We give here the incident polarization characteristics of the three diffraction instruments which have been used:

- On 5C1, we used a Heussler polarizer with incident polarization $p$=0.91, and a cryoflipper with efficiency $e$=0.995(6).

- On 6T2, polarizing bender and adiabatic radiofrequency flipper with $p$=0.981(7) and $e$ >0.99.

- On POLI, $^3$He polarizer was used with varying polarization efficiency 0.90> $p$ >0.75, monitored with time by measuring the $^3$He polarization and adiabatic flipping.

The flipping ratio technique is less sensitive to imperfections of a polarizer, as compared to the experiment with neutron polarization analysis. This is due to the fact that, as seen from Eq. (2) in the main text, the magnetic structure factor is roughly proportional to $p$ and not to the flipping ratio $R_i = \frac{1-p}{1+p}$ of the polarizer. So, the systematic errors due to polarizer imperfections will remain within error bar according to our estimations.

It is well known that extinction effects might have a crucial influence in the polarized neutron data treatment procedure. Therefore, the Becker-Coppens model was used to treat for the extinction in our refinement. We found that the introduction of extinction corrections had no beneficial effect on the refinement. This is likely due to the very small thickness of the crystallites constituting the sample and the rather short neutron wavelengths, 0.84 -1.4 Å, used in the experiment. The same last argument is also valid for the multiple scattering. Next, absorption can introduce error in the flipping ratios only in combination with strong extinction. In the absence of extinction, the flipping ratios do not depend on the absorption. Since in our measurements the extinction was found equal to zero, the absorption correction is not relevant here.

Moreover we have performed in total 7 different experiments with two sample orientations, using three different wavelengths 0.84, 1.18 and 1.4 Å. Hence in each of the experiments the multiple scattering contributions was different. However, the extracted magnetic structures amplitudes were found in agreement within error bars after merging equivalents with the redundancy factor 5.

## III. DATA TREATMENT WITH TWO MAGNETIC FIELD ORIENTATIONS

The flipping ratios were measured at a number of Bragg reflections with two different sample orientations with respect to the applied magnetic field: $H\|[010]$ and $H\|[\bar{1}10]$. The magnetic structure factors at common reflections such as $(0,0,l)$, $(2,0,l)$ and $(2,2,l)$ are shown for each orientation in Fig. **S2**. At large **Q**, Bragg reflection (4,0,0) was measured only with $H\|[010]$ while (4,2,0) and (4,4,0) only with $H\|[\bar{1}10]$.

It is important to notice that the scaling for both orientations is the same and it is consistent to macroscopic SQUID measurements, which show the same saturation magnetic moment for a magnetic field along the two directions. It also supports an isotropic WFM moment, which follows the applied magnetic field[8–10]. We therefore have combined and taken an average of two data sets without an additional scaling.

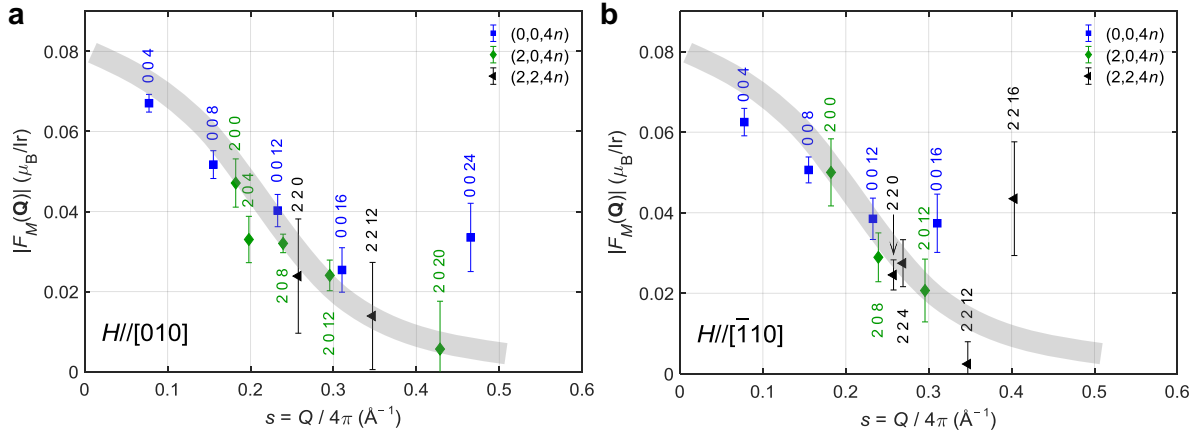

FIG. **S2**. Measured magnetic structure factors with a magnetic field along (a) $H \| [010]$ and (b) $H \| [\bar{1}10]$.

## IV. ANALYSIS USING THE MAXIMUM ENTROPY METHOD

We have tested the maximum entropy method (MEM) by changing the initial positive and negative magnetic moments. The resulting positive and negative magnetic moments are limited by the initial values, thus they are starting points for fitting and also serve as upper bounds. The difference between positive and negative moments was set to the net unit cell magnetic moment, $0.64\mu_\text{B}$/u.c. (8 times $0.08\mu_\text{B}$/Ir). As seen in Fig. **S3**a, the agreement factor defined by $R_w = \sum \left( |F_\text{obs} - F_\text{calc}| / \sigma_F \right) / \sum \left( |F_\text{obs}|/\sigma_F \right)$ is poor ($\approx 15\%$) for only a small negative moment being allowed. It decreases rapidly as the initial negative moment increases, and it is saturated with twice better $R_w \approx 8\%$ for a significant negative moment above $0.25\mu_\text{B}$. Initial values larger than $0.3\mu_\text{B}$, however, do not improve the agreement factor any more and the resulting negative moment does not follow the initial value either. Instead, the both positive and negative densities spread out so the density distribution is only dispersed while the spatial features are unchanged as shown in Fig. **S3**d,g. Therefore, we have chosen $+0.94\mu_\text{B}$ and $-0.30\mu_\text{B}$ as optimized initial positive and negative moments, and the resulting moments are $+0.93\mu_\text{B}$ ($+0.116\mu_\text{B}$/Ir) and $-0.29\mu_\text{B}$ ($-0.036\mu_\text{B}$/Ir), respectively.

In the figure 3a and 3b of the manuscript, the WFM density originates predominantly from the $xy$ orbital, at variance with the standard local $j_\text{eff} = 1/2$ picture. This effect cannot be explained by distortion-induced deviations from the naive equal-weight $j_\text{eff} = 1/2$ state. Indeed, the contribution of the $xy$-orbital to the hole is rather *reduced* by the distortions

19

present in the crystal structure[11,12]. Therefore the $j_{\text{eff}} = 1/2$ description of the hole state cannot account for the present findings, even in its most refined version[13]. We show in Fig. **S4** a schematic illustration of the magnetization for both the local $j_{\text{eff}}=1/2$ model and a predominant $xy$ orbital for the WFM moment. One also remark that positive density lobes are very strongly elongated, in such a way that some magnetization density is delocalized well beyond of the IrO$_6$ octahedra. This is consistent with the very large spatial extent of the $t_{2g}$ orbitals (reaching the nearest neighbouring Ir atoms) found by core-to-core RIXS and *ab initio* calculation[14]. It also could give a support to a direct Ir-Ir exchange mechanism, via electron hopping between the neighboring ions.

## V. ANALYSIS USING A MULTIPOLE EXPANSION MODEL

The magnetization density can be modeled by a superposition of spherical harmonic density functions, known as the *multipolar expansion*[15,16]:

$$m(\mathbf{r}) = \sum_{\text{atoms}} \sum_{l=0}^{\infty} R_n^2(r) \sum_{m=-l}^{l} p_{l,m} y_{l,m}(\theta, \phi), \quad (S3)$$

where $R_n(r)$ are radial wave functions, $y_{l,m}(\theta, \phi)$ are angular density functions, and $p_{l,m}$ are population coefficients. Simple normalized, Slater-type nodeless radial wave functions $R_n(r)$ are defined as:

$$R_n(r) = \sqrt{\frac{(2\zeta)^{2n+1}}{(2n)!}} r^{n-1} \exp(-\zeta r),$$

with $n = 5$ chosen for the $5d$-orbitals. The angular density functions $y_{l,m}(\theta, \phi)$ are the real-valued spherical harmonic functions:

$$y_{l,m}(\theta, \phi) = N_{l,m} P_l^{|m|}(\cos\theta) \begin{cases} \cos m\phi & \text{for } m > 0 \\ \sin|m|\phi & \text{for } m < 0, \end{cases}$$

where $P_l^m$ are the associated Legendre polynomials and $N_{l,m}$ are the normalization factors[17] satisfying

$$\int |y_{l,m}| \, d\Omega = \begin{cases} 1 & \text{for } l = 0 \\ 2 & \text{for } l > 0. \end{cases}$$

Note that these angular *density* functions are distinct from the angular *wave* functions, $Y_l^m$, which are complex-valued and have a different normalization condition, $\int |Y_l^m|^2 \, d\Omega = 1$.

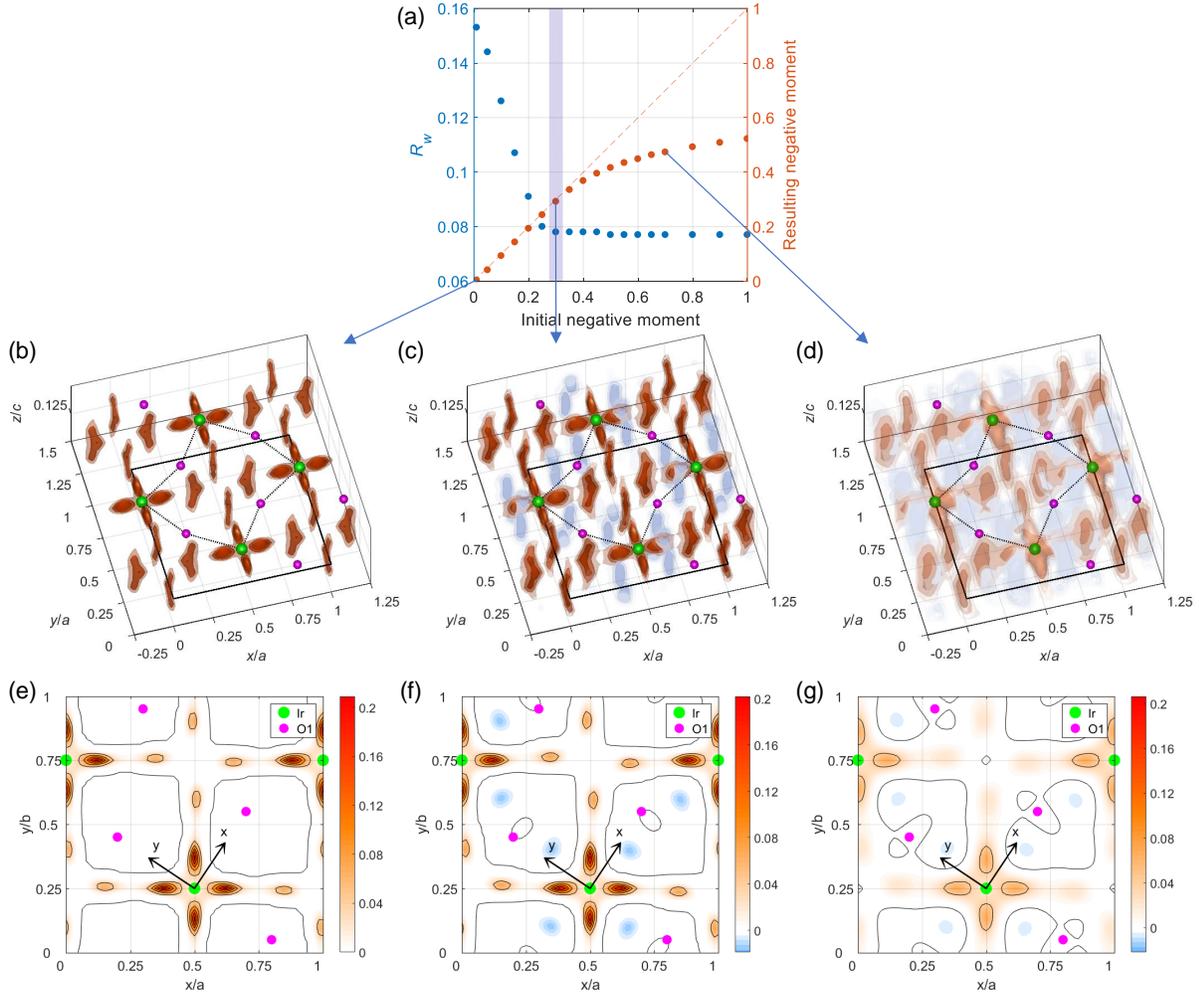

FIG. **S3**. (a) Agreement factor $R_w$ and resulting negative moment per unit cell versus the initial value for negative moment for the MEM analysis. (b-d) The 3D magnetization density distribution on the $z = c/8$ layer reconstructed by the MEM analysis with the initial negative moment (b) $-0.01$, (c) $-0.30$, and (d) $-0.70\mu_\text{B}$ per unit cell. Isosurfaces encompassing 30%, 50% and 70% of the volume density are plotted with a descending opacity according to their isovalues. Red and blue surfaces denote positive and negative magnetizations, respectively. The solid square and dotted lines denote the unit cell and Ir-O bonds, respectively. (e-g) Sliced density contour maps at $(x, y, c/8)$ for the given initial negative moments. The contour step is $0.04\mu_\text{B}/\text{Å}^3$. The blacks arrow correspond to the Ir-O bonding directions tilted by 11° from the crystallogaphic tetragonal axes.

The magnetic form factors corresponding to the model magnetization density in Eq. (S3)

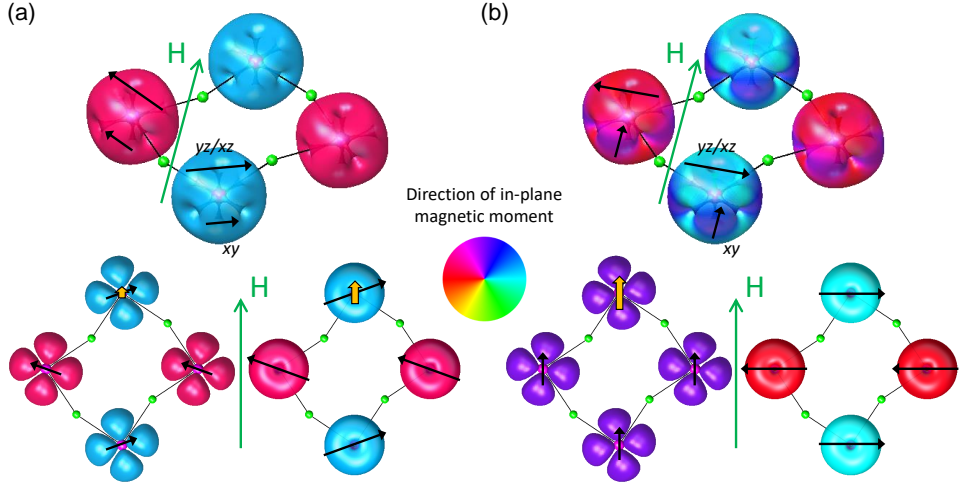

FIG. S4. (a) Magnetization density distribution for the local $j_{\text{eff}}=1/2$ picture. The $xy$, $yz$ and $xz$ magnetic components have the same pseudospin moment (black arrows) at each Ir site. The magnetic moments at adjacent sites are aligned antiferromagnetically, perpendicular to the applied magentic field H, and the canted FM moment (orange arrows) along $H$ appears in both $xy$ and $yz/xz$ components. (b) Magnetization density distribution for the proposed $k$-dependent $j_{\text{eff}}=1/2$ picture. The magnetic moment in the $xy$ component is aligned along H, while one in the $yz/xz$ component is perfectly antiparallel and perpendicular to $H$. The WFM moment solely comes from the $xy$ component.

become

$$f(\mathbf{Q}) = \sum_{l=0}^{\infty} \phi_l(Q) \sum_{m=-l}^{l} p_{l,m} y_{l,m}(\theta_k, \phi_k)$$

with

$$\phi_l(Q) = 4\pi i^l \int_0^\infty R_n^2(r) j_l(Qr) r^2 dr,$$

where $j_l$ is a spherical Bessel function of order $l$. Then the least-squares refinement with the measured $F_M$ was done using the MPLSQ program[18]. For the refinement, the total magnetic moment is constrained: $\mu_{\text{Ir}} + 2\mu_{\text{O1}} + 2\mu_{\text{O2}} = 0.08\mu_B/\text{Ir}$.

The multipole expansion in general can describe any atomic magnetization or charge density when all $l$ and $m$ values are allowed in the expansion. However, we used a limited number of harmonic functions ($d_{00}$, $d_{20}$, $d_{40}$, $d_{44+}$) that can describe the $d$-orbital density and are allowed by the site symmetry, $D_{4h}$. There is also a direct relation matrix between multipole functions and orbital populations. The multipole density functions allowed by

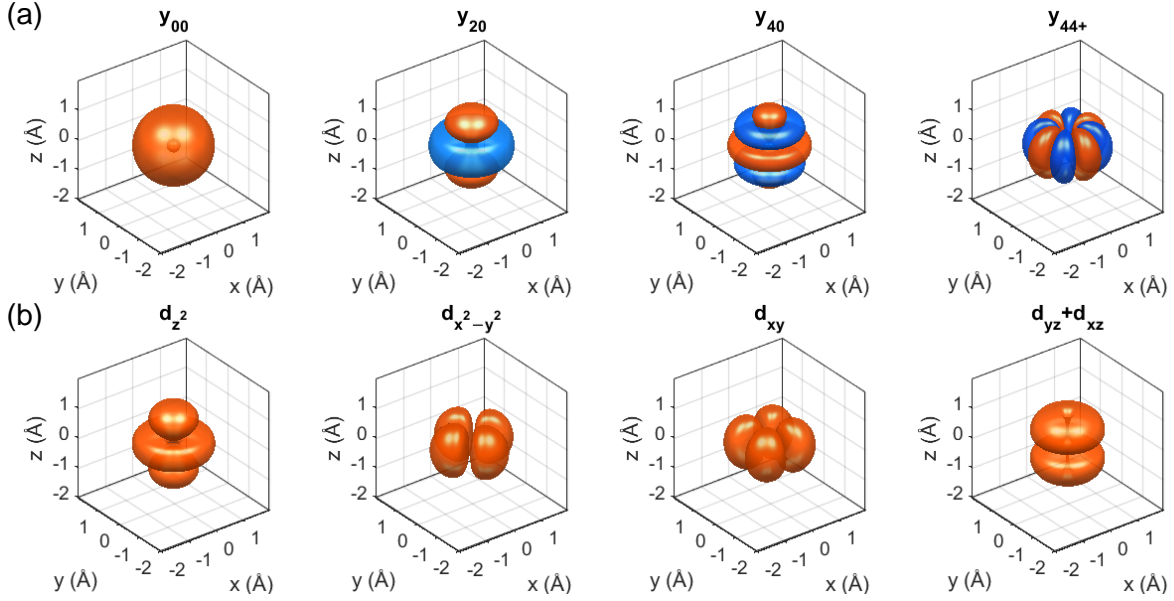

FIG. **S5**. (a) Multipole density functions allowed by $4/mmm$ symmetry. Red and blue surface denote positive and negative density, respectively. (b) Density distribution of $d$ orbitals constructed by a linear combination of multipole density functions.

the point group $D_{4h}$ ($4/mmm$) are shown in Fig. **S5**a. Because the spherical harmonic functions constitute a complete set in the spherical harmonic point group, the orbital density distribution given by a square of wave function $|Y_l^m|^2$ must be a linear combination of spherical harmonic density functions $y_{l,m}$[16]. The orbital-multipole relations for the point group $D_{4h}$ are defined as $p_l^m |Y_l^m|^2 = p_{i,j} y_{i,j}$ with

$$\begin{pmatrix} p_2^0 \\ p_2^{\pm 1} \\ p_2^2 \\ p_2^{-2} \end{pmatrix} = \begin{pmatrix} 0.200 & 1.039 & 1.396 & 0.000 \\ 0.200 & 0.520 & -0.931 & 0.000 \\ 0.200 & -1.039 & 0.233 & 1.570 \\ 0.200 & -1.039 & 0.233 & -1.570 \end{pmatrix} \begin{pmatrix} p_{0,0} \\ p_{2,0} \\ p_{4,0} \\ p_{4,4} \end{pmatrix}.$$

The $d$ orbital densities reconstructed using the spherical density functions are also shown in Fig. **S5**b. Using this orbital-multipole relations, the orbital populations were obtained from the fitted density function populations.

In order to describe the large spatial extent of the positive magnetization density, the radial wave function was also examined. It is clear as shown in Fig. **S6**f that the theoretical radial function with the radial exponent $\zeta = 3.74$ is not fit well ($R_w \approx 22\%$) with the measured magnetic structure factor and also not able to describe the wide spatial extent

23

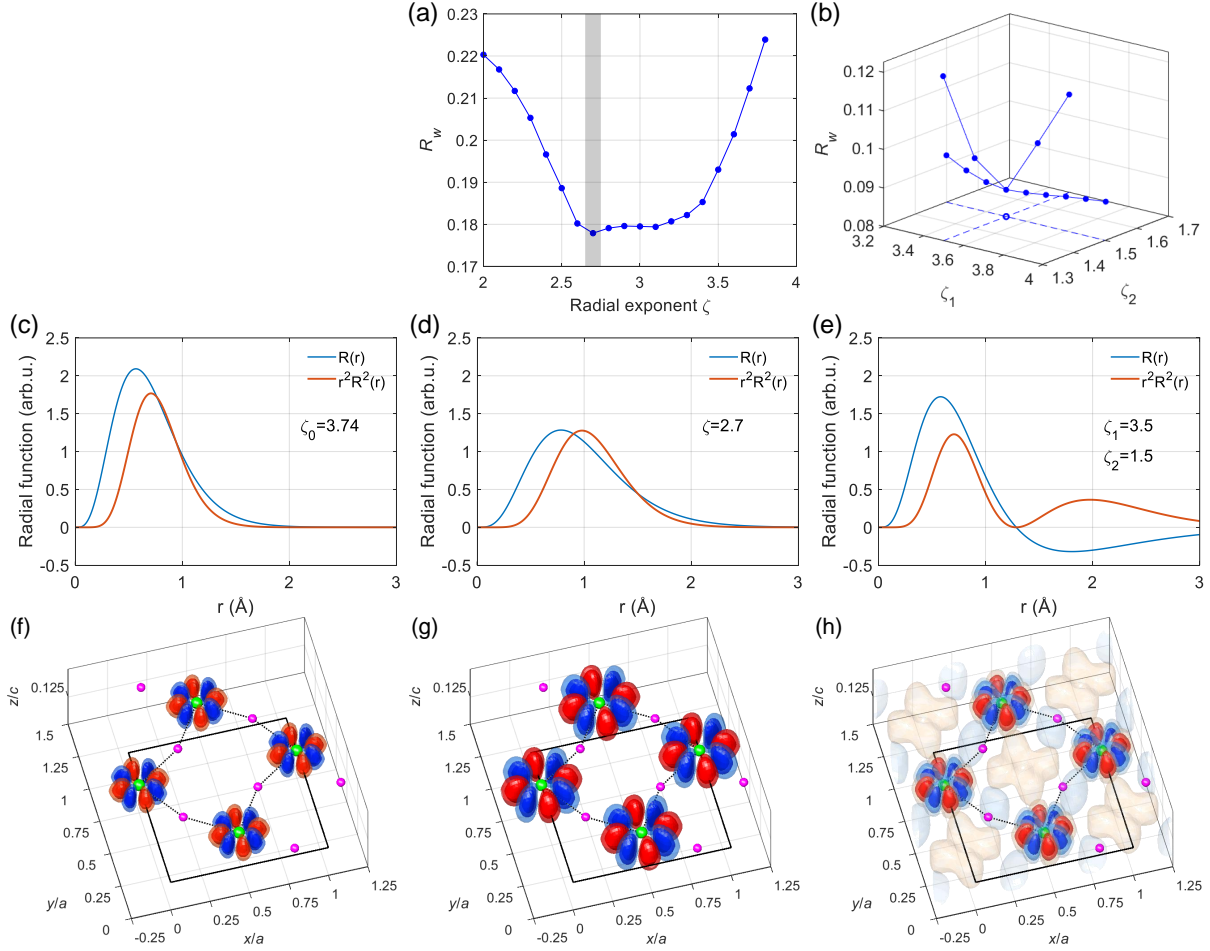

FIG. **S6**. (a) Agreement factor $R_w$ versus the radial exponent $\zeta$ for the multipole expansion model with a single radial exponent. The agreement factor decreases by ∼20% at $\zeta \simeq 2.7$ compared with the theoretical radial exponent $\zeta_0 = 3.74$. (b) Agreement factor $R_w$ with two radial exponents $\zeta_1$ and $\zeta_2$. Radial wave functions $R(r)$ and radial distribution functions $r^2 R^2(r)$ with (c) the theoretical radial exponent for isolated Ir atoms, (d) an optimized single radial exponent, and (e) double radial exponents. (f-h) Reconstructed magnetization density distributions with different radial wave functions in (c-e).

found by the MEM analysis. The best value $\zeta = 2.7$ for a single radial exponent was also refined, but the agreement factor was not improved significantly ($R_w \approx 18\%$) as shown in Fig. **S6**a. Therefore, we introduced an additional radial exponent. The radial wave function with two radial exponents provides significantly better agreement ($R_w \approx 9\%$) with measured magnetization density and describes the large extent of observed magnetization density as

24

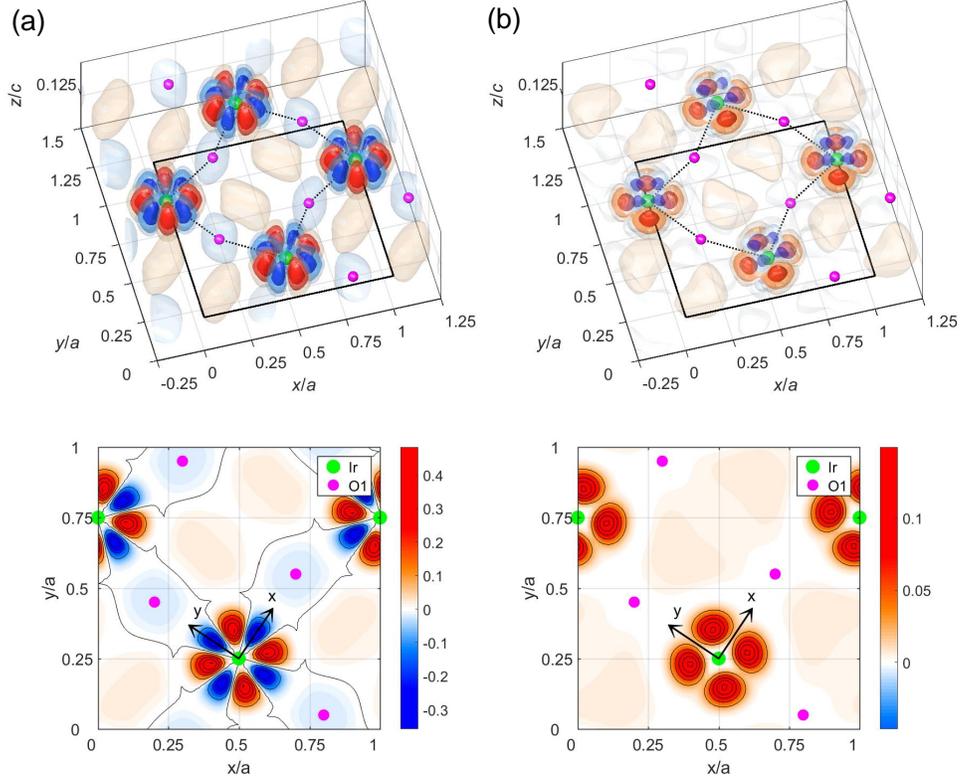

FIG. **S7**. Magnetization density distribution obtained by orbital density model fit with (a) all five $d$ orbitals and (b) only three $t_{2g}$ orbitals.

shown in Fig. **S6**h.

Further, between the positive lobes, negative density lobes occurs as well which are more pronounced in the multipole refinement than in the MEM results (see the Fig. 3 of the main mansucript). They are about 60% of positive ones and surprisingly have $d_{x^2-y^2}$ symmetry, requiring an admixing of the $e_g$ orbital to the ground state. Ii is worth to remind that the MEM method is model-free whereas in the multipole expansion method, we have necessarily to start with certain multipoles corresponding to the $d$-orbitals. The $d_{x^2-y^2}$ contribution could be overestimated due to insufficient modeling in the multipole description. However, the main result in the MEM - dominant $xy$ and no $yz/xz$ character - is well reproduced by the multipole expansion method. We tried various models (see Fig. **S6**), and the given result is still the best fit.

In order to examine a possible magnetization density expected from a strong $d$-$p$ hybridization at the oxygen site, the refinements with a spherical magnetization density at

the O1 and O2 sites were also performed, but contrary to expectations, no evidence for the existence of the oxygen moment was found. The agreement factor did not become better and also the refined magnetization density at the O sites is statistically negligible.

We have also examined to fit the data with actual orbital density functions instead of multipole density functions. When all possible $5d$ orbitals are allowed, it obviously gives the same result ($R_w \approx 9\%$) with similar population of each orbitals as shown in Fig. **S7**a. In addition, as it is known to consist mostly of $t_{2g}$ orbitals, the model fit allowing only $d_{xy}$, $d_{yz}$, $d_{xz}$ orbitals was also performed. As shown in Fig. **S7**b, the negative magnetic density with $d_{x^2-y^2}$ symmetry disappears as expected, but the agreement becomes twice worse ($R_w \approx 17\%$). Therefore, we conclude that the observed magnetic octapolar shape, mainly from $y_{44+}$ angular density function, is essential to describe our measured data.

---



\end{document}